\documentclass[]{spie}  %>>> use for US letter paper
\usepackage{amsmath,amsfonts,amssymb}
\usepackage{graphicx}
\usepackage[colorlinks=true, allcolors=blue]{hyperref}

\title{Maximising the mid-infrared high-contrast performance of ELT/METIS despite water vapour seeing}

\author[a]{Olivier Absil}
\author[a]{Gilles Orban de Xivry}
\author[b]{Prashant Pathak}
\author[b]{Abhirami S. Raghu}
\author[c]{Benjamin Courtney-Barrer}
\author[d]{Roy van Boekel}
\author[d]{Thomas Bertram}
\author[d]{Markus Feldt}
\author[e]{Matthew Kenworthy}
\author[e]{Bernhard Brandl}
\affil[a]{STAR Institute, Universit\'e de  Li\`ege, all\'ee du Six Ao\^ut 19c, 4000 Li\`ege, Belgium}
\affil[b]{Department of SPASE, Indian Institute of Technology Kanpur, 208016, Uttar Pradesh, India}
\affil[c]{Australian Astronomical Optics \& School of Mathematical and Physical Sciences, Macquarie University, Sydney 2109, Australia}
\affil[d]{Max Planck Institute for Astronomy, D-69117 Königstuhl 17, Heidelberg, Germany}
\affil[e]{Leiden Observatory, Leiden University, Postbus 9513, 2300 RA Leiden, The Netherlands}

\authorinfo{Further author information: (Send correspondence to O.A.)\\O.A.: E-mail: olivier.absil@uliege.be}

\begin{document} 
\maketitle

\begin{abstract}
The Mid-infrared ELT Imager and Spectrograph (METIS) will be equipped with a SCAO system delivering Strehl ratios above 90\% at L band (3.5 - 4.1~\textmu m) and close to 99\% at N band (8 - 13~\textmu m) on bright stars. Yet, the actual wavefront quality seen by the METIS coronagraphic modules used for high-contrast imaging will be significantly affected by water vapour seeing, which add a strong chromatic component to dry air seeing in the mid-infrared. We analysed two years of VLTI/GRAVITY fringe tracker archives to assess the variability of differential water vapour column density at millisecond timescales on ELT scales. Our analysis suggests that water vapour seeing will add a median wavefront error of about 175~nm rms at N band, consisting mostly of low-order aberrations, with around 150~nm rms of tip-tilt errors. If not corrected, this effect would degrade the achievable sensitivity limits in terms of contrast by more than two magnitudes. To mitigate this effect, we plan to deploy a focal-plane wavefront sensing and control algorithm based on an asymmetric pupil using a deep learning approach. After briefly discussing the practical impacts of focal-plane wavefront control in METIS, we compare the expected high-contrast imaging performance with and without focal-plane wavefront control.

\end{abstract}

% Include a list of keywords after the abstract 
\keywords{extremely large telescope, high-contrast imaging, thermal infrared, atmospheric turbulence, water vapour, end-to-end simulations}

\section{INTRODUCTION}
\label{sec:intro}  % \label{} allows reference to this section

In a previous work\cite{Absil22}, we have shown that the high-contrast imaging (HCI) performance of METIS, the ELT mid-infrared imager and spectrograph, could be significantly affected by the effect of turbulence in the water vapour content of the atmosphere. Also referred to as water vapour seeing (WVS), this phenomenon creates strong chromatic wavefront errors in the mid-infrared domain. We refer the reader to that previous work for a full description of the WVS phenomenon, and how its strength can be measured with the GRAVITY fringe tracker of the Very Large Telescope Interferometer (VLTI) through the chromaticity of the phase delay relative to the group delay at K band, which can be translated into a measurement of the differential column density of water vapour above the VLTI telescopes. The time variability of this quantity is directly linked to the additional wavefront errors induced by water vapour variability in the atmosphere. When expressed as an equivalent quantity of (precipitable) water vapour in units of \textmu m, this quantity, noted $\sigma_{\rm WV}$, is what we define as WVS (even though, stricto sensu, seeing is supposed to refer to an angle).

\begin{figure}[t]
 \begin{center}
  \begin{tabular}{c} %% tabular useful for creating an array of images 
   \includegraphics[width=0.48\textwidth]{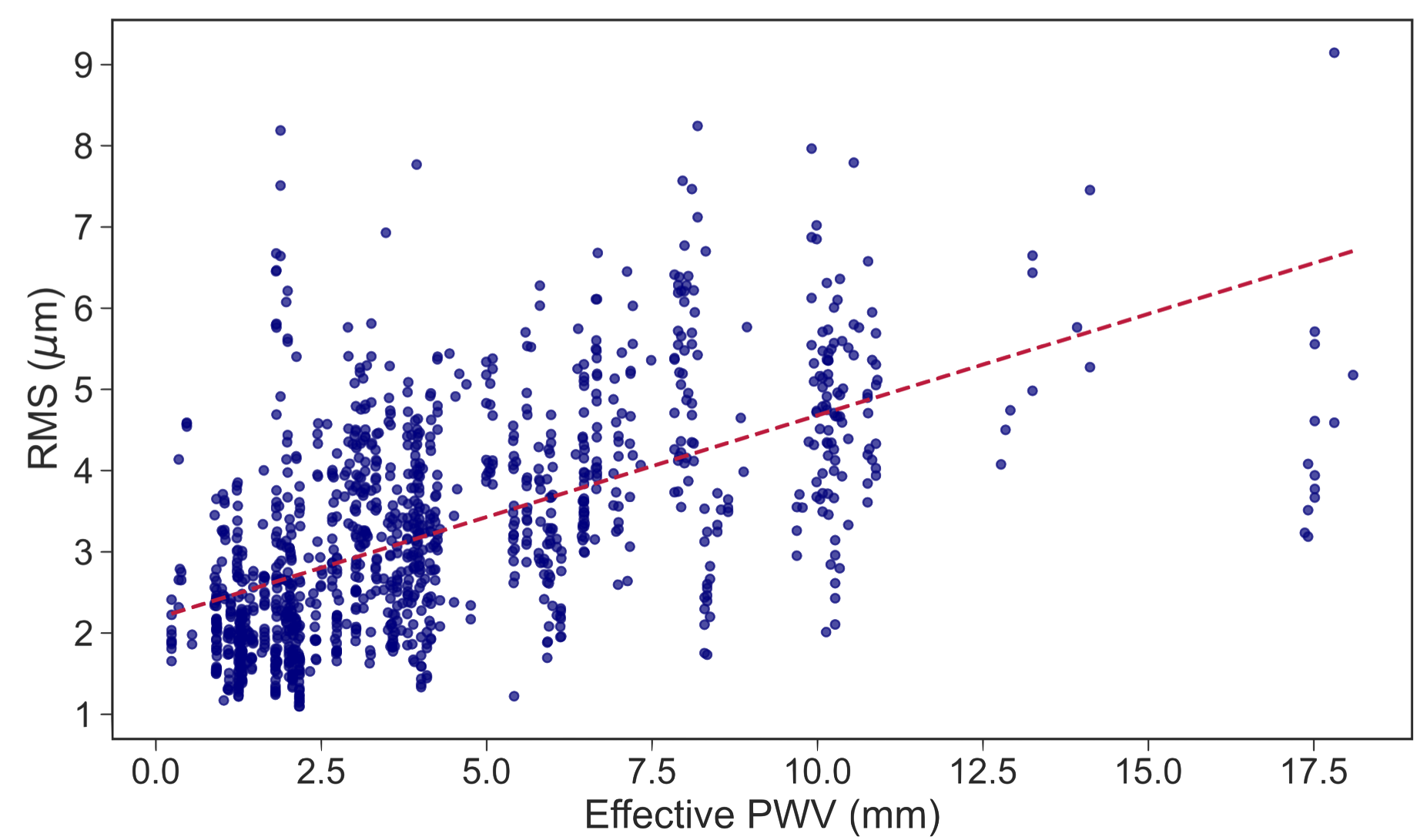} \hspace*{2mm}
   \includegraphics[width=0.47\textwidth]{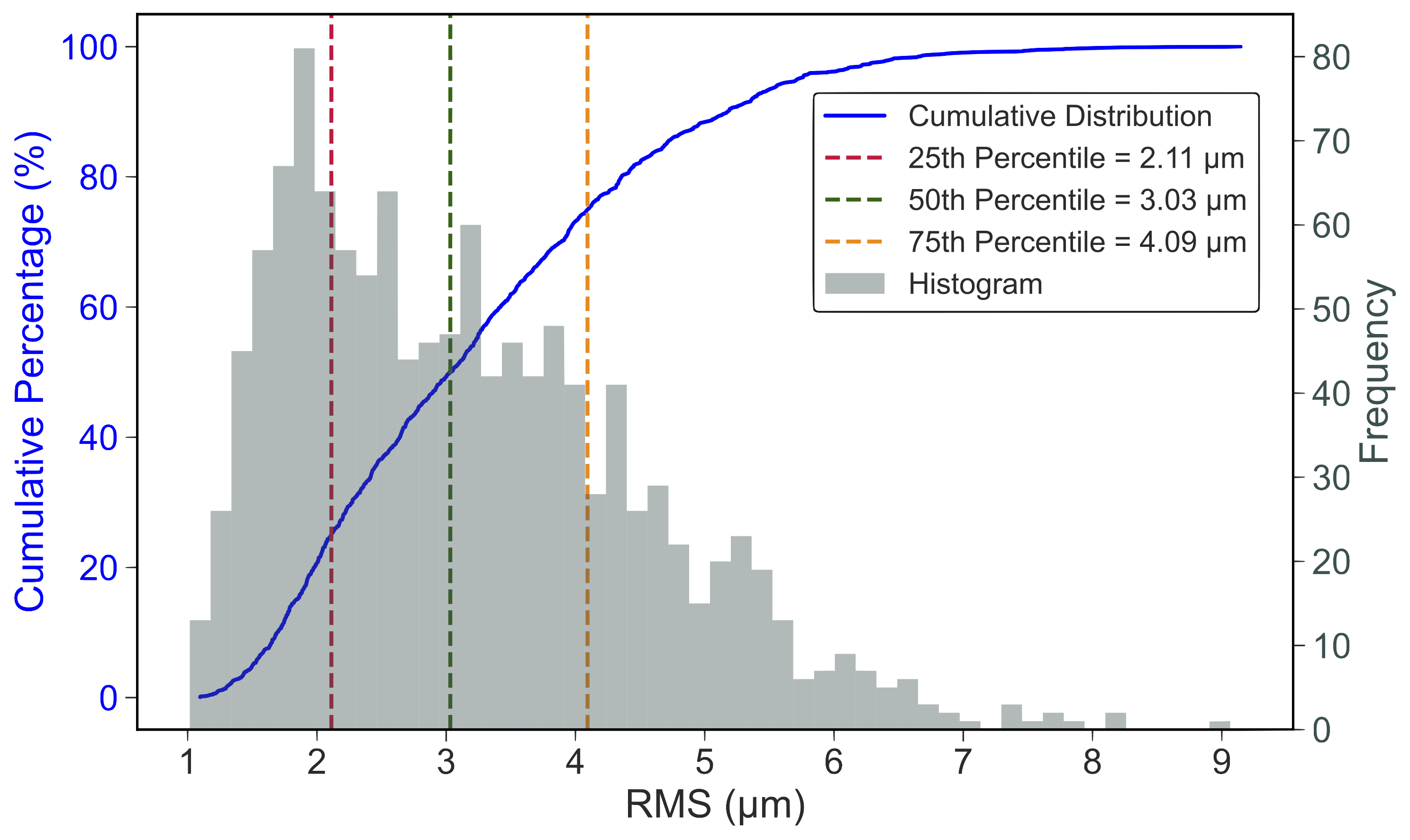}
  \end{tabular}
 \end{center}
 \caption[example] 
%>>>> use \label inside caption to get Fig. number with \ref{}
 { \label{fig:WVS} \textit{Left.} Water vapour seeing (WVS) measurements on the short baselines of the VLTI auxiliary telescope array (baseline lengths similar to the ELT pupil size), as a function of the precipitable water vapour (PWV) content in the atmosphere. The red line shows the best-fit linear trend, although this should not be considered as a valid model to predict WVS from PWV. \textit{Right.} Histogram and cumulative distribution of WVS ($\sigma_{\rm WV}$). The grey bars represent the frequency distribution of $\sigma_{\rm WV}$, while the blue curve shows the cumulative percentage. The dashed lines indicate the 25th (red), 50th (median, green), and 75th (orange) percentiles. Both plots are from Ref.~\citenum{Raghu26}.}
\end{figure}

Our initial estimate of the HCI performance degradation presented in Ref.~\citenum{Absil22} was based on a single measurement of the strength of WVS at Cerro Paranal, which significantly limited its scope of validity. Since then, we have carried out a much more comprehensive re-analysis of more than one year of GRAVITY fringe tracker data. This analysis is presented in a paper \cite{Raghu26} whose main results are summarised here for convenience. One of the main results from this analysis is that the strength of WVS increases with the amount of precipitable water vapour (PWV) in the atmosphere, but (somewhat counter-intuitively) does not seem to depend on the strength of dry air seeing. The relationship between WVS and PWV is not found to be a 1:1 proportional relation though, as shown in Figure~\ref{fig:WVS} (left). The large variability around the linear trend in that figure is understood to be mostly due to noise in the GRAVITY fringe tracker measurements (see Ref.~\citenum{Raghu26} for details). Another important result from this analysis was to derive the statistics of WVS. The histogram of WVS and its cumulative distribution are illustrated in Figure~\ref{fig:WVS} (right). The boundaries between quartiles (i.e., percentiles 25, 50 and 75) are illustrated in the figure, with the associated WVS level in \textmu m rms. Here, we further convert this histogram into ESO-compliant quartile definitions, where the quartile conditions are defined as the median condition within each of the quartiles (i.e., 12.5\%, 37.5\%, 62.5\% and 87.5\% for Q1 to Q4, respectively). These quartile conditions are provided in Table~\ref{tab:WVS}, where we also give their conversion in terms of differential water vapour column density ($\sigma_{\Sigma}$) in cm$^{-2}$, with 1 \textmu m of precipitable water vapour corresponding to a column density of $0.33 \times 10^{19}$~cm$^{-2}$, owing to the density and molecular mass of water.

\section{CHROMATIC WAVEFRONT ERRORS AT METIS WAVELENGTHS}
\label{sec:WFE}  % \label{} allows reference to this section

Converting the rms column density $\sigma_{\Sigma}$ [cm$^{-2}$] into an rms OPD is just a matter of multiplying this rms column density by the appropriate refractive index. For this, we use the concept of ``water vapour displacing air'' (WDA), where fluctuations in the WV column density, under constant pressure and temperature, lead to opposite variations in the dry air column density. The refractivity of WDA ($n_{\rm WDA}-1$) at infrared wavelengths is based on the refractive index of dry air and water vapour,\cite{Mathar04} adapted to the temperature and pressure conditions of Cerro Paranal. We obtained through Jeff Meisner (personal communication, see also Ref.~\citenum{Meisner03}) a tabulated version of the normalised refractive index of air and water vapour described in Ref.~\citenum{Mathar04}, expressed in fs/(mol/m$^2$). These units allow to focus on the properties of a gas without specifying temperature, pressure, or concentration in a mixture. We refer the reader to our previous analysis of WVS (Ref.~\citenum{Absil22}) for more details on these concepts and for illustrations of the refractive indices of air, water vapour, and WDA. For a more convenient use in the following, we convert the refractivity of WDA into units of cm$^3$/molecule by multiplying the tabulated values in fs/(mol/m$^2$) by the speed of light and dividing by the Avogadro number. This quantity, noted $N_{\rm WDA}$ and sometimes referred to as specific or molecular refractivity, takes the following values in four representative METIS filters, relative to the K band where the METIS single-conjugate adaptive optics (SCAO) system operate:
\begin{itemize}
    \item HCI-L long (3.8 \textmu m): $N_{\rm WDA}(L - K) = 2.42 \times 10^{-25}$ cm$^3$ per molecule
    \item CO ref (4.8 \textmu m): $N_{\rm WDA}(M - K) = 7.04 \times 10^{-25}$ cm$^3$ per molecule
    \item N1 (8.6 \textmu m): $N_{\rm WDA}(N1 - K) = 1.34 \times 10^{-24}$ cm$^3$ per molecule
    \item N2 (11.2 \textmu m): $N_{\rm WDA}(N2 - K) = 3.02 \times 10^{-24}$ cm$^3$ per molecule
\end{itemize}
The new units allow us to obtain the rms OPD associated with a given rms water vapour column density variability ($\sigma_{\Sigma}$) by a simple multiplication with the specific refractivity. The associated rms OPDs, converted into nm, are given in Table~\ref{tab:WVS}.

\begin{table}[t]
\caption{Water vapour quartiles expressed in terms of equivalent precipitable water vapour variability ($\sigma_{\rm WV}$), water vapour column density variability ($\sigma_{\Sigma}$), and inferred temporal OPD variation in nm rms on the VLTI compact AT baselines in four representative METIS HCI filters relative to the centre of K band, for the four quartiles (following the ESO definition) and median conditions.} 
\label{tab:WVS}
\begin{center}       
\begin{tabular}{|c|ccccc|} 
\hline
\rule[-1ex]{0pt}{3.5ex}   & Q1 & Q2 & med & Q3 & Q4 \\
\hline
\rule[-1ex]{0pt}{3.5ex}  $\sigma_{\rm WV}$ [\textmu m] & 1.76 & 2.54 & 3.03 & 3.53 & 4.91    \\
\hline
\rule[-1ex]{0pt}{3.5ex}  $\sigma_{\Sigma}$ [$10^{19}$~cm$^{-2}$] & 0.59 & 0.85 & 1.01 & 1.18 & 1.64 \\
\hline
\rule[-1ex]{0pt}{3.5ex}  $\sigma_{\rm OPD}$(HCI-L~long) [nm] & 14 & 21 & 24 & 29 & 40 \\
\hline
\rule[-1ex]{0pt}{3.5ex}  $\sigma_{\rm OPD}$(CO~ref) [nm] & 42 & 60 & 71 & 83 & 115  \\
\hline
\rule[-1ex]{0pt}{3.5ex}  $\sigma_{\rm OPD}$(N1) [nm] & 79 & 114 & 136 & 158 & 220 \\
\hline
\rule[-1ex]{0pt}{3.5ex}  $\sigma_{\rm OPD}$(N2) [nm] & 178 & 257 & 305 & 356 & 495 \\
\hline
\end{tabular}
\end{center}
\end{table} 

Armed with the temporal rms OPDs listed in Table~\ref{tab:WVS}, we can produce a realistic time series of phase screens for WVS. This is as simple as generating a sequence of turbulent phase screens with an appropriate statistical behaviour, and scaling this sequence so that the rms OPD extracted on baselines similar to the VLTI-AT compact configuration matches the values given in Table~\ref{tab:WVS}. We remind the reader of the excellent match between the VLTI-AT compact configuration and the ELT pupil, which was already illustrated in Figure~6 of Ref.~\citenum{Absil22}, making this scaling even more relevant to the case of METIS. As discussed in Ref.~\citenum{Absil22}, we assume that WVS follows a von Karman turbulence distribution, with an outer scale of 500~m. As already described in Ref.~\citenum{Absil22}, we generate a 1-min sequence of Kolmogorov-von Karman phase screen and sample it on the six VLTI-AT baselines, resulting in six time series of piston. Over the 1-min sequence, the average of the six rms temporal OPD series is then 8814~nm. We divide all phase screens by this normalising factor, and multiply by one of the values listed in Table~\ref{tab:WVS} to obtain a properly scaled sequence of phase screens, corresponding to WVS at the chosen wavelength and quartile.

While the phase screen sequences generated here are all we need to feed METIS end-to-end simulations and/or METIS focal-plane wavefront control simulations, it is useful to convert the values given in Table~\ref{tab:WVS} into associated spatial wavefront errors, so that they can be compared to other contributors such as static NCPA, chromatic beam wander,\cite{Bone26} or SCAO residuals\cite{Feldt24}. The median rms wavefront errors for the various phase screen sequences generated above are given in Table~\ref{tab:WFE}. The conversion factor from temporal rms on VLTI-AT baselines to spatial rms at ELT scale is 0.573 for these specific assumptions in terms of turbulence distribution.

\section{CONSEQUENCES ON METIS-HCI OPERATIONS}
\label{sec:METIS}  % \label{} allows reference to this section

The additional wavefront errors induced by WVS in the red part of the N band are well beyond the expected residual wavefront errors provided by SCAO in closed loop, which amount to about 128 nm rms\cite{Feldt24}. This means that the N-band HCI performance will be driven by WVS rather than SCAO residuals, and justifies the need for real-time closed-loop control of WVS (and hence of NCPA in general). Our strategy to implement this control relies on focal-plane wavefront sensing using the data stream from the imaging cameras inside the METIS LM-band or N-band imager (IMG-LM and IMG-N). In order to allow for instantaneous phase retrieval from the IMG camera data stream, we use an asymmetric Lyot stop downstream of the vortex coronagraph, and train a deep neural network to map the recorded intensity pattern with the pupil-plane phase aberrations, as described in Ref.~\citenum{Orban24}. We refer the reader to that paper for a complete description of the NCPA measurement and control strategy, as well as the expected NCPA control performance. Here, we describe the practical consequences of WVS and its real-time control on METIS operations, and explore the gain in HCI performance provided by focal-plane wavefront control in the specific case of alpha Centauri observations at N band.

\begin{table}[t]
\caption{Wavefront error due to WVS in nm rms at the central wavelength of four representative METIS HCI filters, for the quartiles and median conditions defined in Table~\ref{tab:WVS}, assuming SCAO wavefront control at K band.} 
\label{tab:WFE}
\begin{center}       
\begin{tabular}{|c|ccccc|} 
\hline
\rule[-1ex]{0pt}{3.5ex}   & Q1 & Q2 & med & Q3 & Q4 \\
\hline
\rule[-1ex]{0pt}{3.5ex}  HCI-L long (3.8~\textmu m) & 8.2 & 11.8 & 14.0 & 16.4 & 22.8 \\
\hline
\rule[-1ex]{0pt}{3.5ex}  CO ref (4.8~\textmu m) & 23.8 & 34.3 & 40.8 & 47.7 & 66.3  \\
\hline
\rule[-1ex]{0pt}{3.5ex}  N1  (8.6~\textmu m) & 45.5 & 65.5 & 77.8 & 91.0 & 126 \\
\hline
\rule[-1ex]{0pt}{3.5ex}  N2 (11.2~\textmu m) & 102 & 147 & 175 & 205 & 284 \\
\hline
\end{tabular}
\end{center}
\end{table} 

\subsection{Consequences on SCAO operations}

The main consequence on SCAO operations is the need to accommodate NCPA corrections in SCAO closed loop operations at a level that corresponds to the input perturbation. As discussed in Ref.~\citenum{Feldt24}, this will be implemented through a combination of modulator offsets (for tip-tilt corrections) and slope offsets in the pyramid wavefront sensor (for higher-order modes). Compensating for NCPA means that the SCAO will see a significant amount of aberration at the tip of the pyramid WFS. This is expected to reduce the SCAO performance, and thereby increase the rms of residual phase errors.

To properly design and test the described WVS correction strategy in the SCAO module, it is useful to separate the WVS wavefront error into tip-tilt from higher order modes, as they are expected to be corrected by different mechanisms (modulator offsets for tip-tilt vs pyramid slope offsets for higher-order modes). In his seminal paper, Noll\cite{Noll76} found that tip-tilt contains about 87\% of the wavefront variance for Kolmogorov turbulence. However, in presence of a finite outer scale (von Karman turbulence spectrum), the relative strength of tip-tilt decreases.\cite{Winker91} Assuming an outer scale of 500~m for water vapour turbulence (see above), the relative contribution of tip-tilt to the wavefront variance decreases to about 71\%. In Table~\ref{tab:SCAO}, we provide the estimated contribution of tip-tilt vs higher order modes to the wavefront errors at various wavelengths in median conditions.

The NCPA correction strategy will need to cope with wavefront errors significantly larger than the median errors reported in Table~\ref{tab:SCAO}. A reasonable assumption to make here is that the NCPA correction needs to be robust up to about three times the rms values reported in Table~\ref{tab:SCAO}, which would mean being able to implement tip-tilt offsets up to 10~mas with the SCAO modulator, and wavefront set-point offsets up to almost 300~nm rms for the pyramid wavefront sensor. In Ref.~\citenum{Feldt24}, we show that SCAO operations are robust to pyramid slope offsets up to 300~nm rms, with a Strehl degradation of less then one percent at L band (or even less at N band) due to the reduced SCAO performance. This validates, at least in simulation, our NCPA control strategy.

\begin{table}[t]
\caption{Relative contribution of tip-tilt and higher order errors to the median WVS-induced wavefront errors described in Table~\ref{tab:WFE}.} 
\label{tab:SCAO}
\begin{center}       
\begin{tabular}{|c|c|cc|c|} 
\hline
\rule[-1ex]{0pt}{3.5ex}   & Median WFE & \multicolumn{2}{|c|}{Tip-tilt (WFE and angle)} & Higher orders \\
\hline
\rule[-1ex]{0pt}{3.5ex}  HCI-L long (3.8~\textmu m) & 14.0 nm & 11.8 nm & 0.26 mas & 8.1 nm \\
\hline
\rule[-1ex]{0pt}{3.5ex}  CO ref (4.8~\textmu m) & 40.8 nm & 34.4 nm & 0.77 mas & 22.0 nm \\
\hline
\rule[-1ex]{0pt}{3.5ex}  N1  (8.6~\textmu m) & 77.8 nm & 65.6 nm & 1.46 mas & 41.9 nm \\
\hline
\rule[-1ex]{0pt}{3.5ex}  N2 (11.2~\textmu m) & 175 nm & 147.5 nm & 3.29 mas & 94.2 nm \\
\hline
\end{tabular}
\end{center}
\end{table}

\subsection{Consequences on HCI operations}

The main consequence of this study on HCI operations is that HCI observations at N band would benefit from the best possible conditions in terms of WVS, as it will make the implementation of NCPA corrections easier in the SCAO module (smaller slope offsets), and will lead to the smallest possible wavefront errors at the level of the coronagraph after NCPA closed-loop correction. In Ref.~\citenum{Raghu26}, we explored possible correlations between WVS and various other atmospheric parameters, and only identified a correlation with PWV, illustrated in Figure~\ref{fig:WVS} (left). Because of the large scatter of WVS around the linear trend shown that figure, it is hard to come up with a clear recommendation in terms of maximum PWV level for N-band HCI observations. PWV has been monitored at Cerro Paranal for more than 10 years with LHATPRO\cite{Sarazin13}. The resulting statistics can be found on the ESO astro-climatology web page\footnote{\url{https://www.eso.org/sci/facilities/paranal/astroclimate/climatology/lhatpro.html}}, with the night-time 25\%, 50\% and 75\% percentiles respectively around 1.5, 2.5 and 4~mm (see also Ref.~\citenum{Sarazin13}). To ensure that WVS is at least better than the Q3 conditions listed in Tables~\ref{tab:WVS} and \ref{tab:WFE} (and generally better than median) during N-band HCI observations, Figure~\ref{fig:WVS} (left) suggests that a criterion of 1.5~mm on the PWV would be reasonable. This would also provide the lowest sky background levels, as sky background is known to correlate with PWV level as well\cite{Pathak22}.

At LM bands, the contribution of WVS is sufficiently small that the WVS conditions should not be a decisive criterion to schedule HCI observations, even though one would still generally benefit from lower PWV for various reasons (sky transparency, strength of atmospheric refraction, etc.).

\subsection{Consequences on HCI performance}

\begin{figure}[t]
 \begin{center}
  \begin{tabular}{c} %% tabular useful for creating an array of images 
   \includegraphics[width=0.8\textwidth]{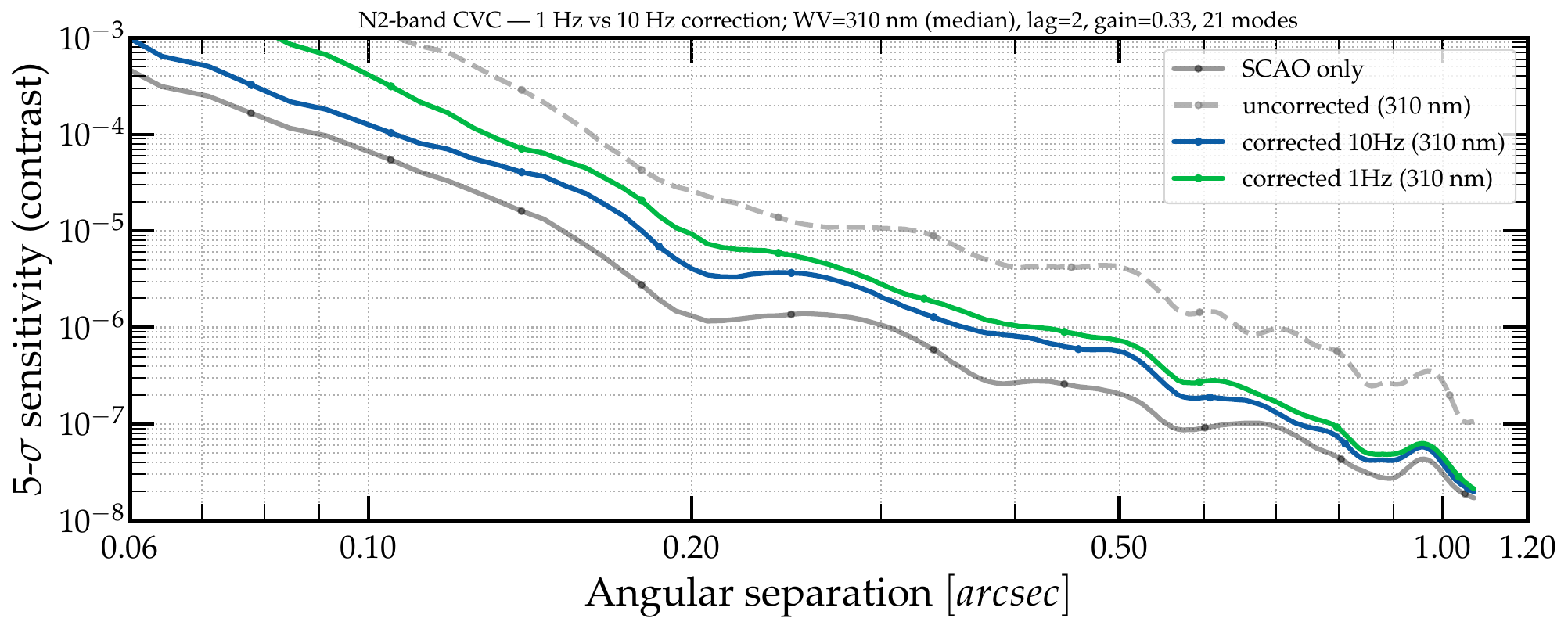}
  \end{tabular}
 \end{center}
 \caption[example] 
%>>>> use \label inside caption to get Fig. number with \ref{}
 { \label{fig:ADI} Post-ADI performance of the METIS classical vortex coronagraph (CVC) in the N2 filter in presence of WVS, either uncorrected (dashed grey) or with 20 Zernike modes corrected at 1 Hz (green) or 10 Hz (blue), in median WVS conditions (305 nm of OPD, i.e., 175 nm WFE). The SCAO-limited performance is shown in solid grey for reference.}
\end{figure} 

Here we illustrate the effect of WVS on HCI performance at N band, where it has the largest effect, and how closed-loop control based on focal-plane wavefront sensing can help mitigate this effect. A first important question to address is the speed at which focal-plane wavefront control should operate. In Fig.~\ref{fig:ADI}, we show the influence of closed-loop focal-plane wavefront control on N-band performance for our standard case of a 1-h ADI sequence on a star at declination $-5^{\circ}$ around meridian crossing (as in Ref.~\citenum{Carlomagno20}), with no WVS control (dashed grey), and control of the first 20 Zernike modes of WVS at a loop frequency of 1~Hz (green) and 10~Hz (blue), in median WVS seeing conditions. This figure illustrates the huge HCI performance gain enabled by focal-plane wavefront control, with post-ADI sensitivity improving typically by a factor 5 with respect to the case where WVS is not corrected. It also illustrates the gain of operating the loop at 10 Hz rather than 1 Hz, with an improvement by a factor up to 2-3 at short angular separations.

\begin{figure}[t]
 \begin{center}
  \begin{tabular}{c} %% tabular useful for creating an array of images 
   \includegraphics[width=0.8\textwidth]{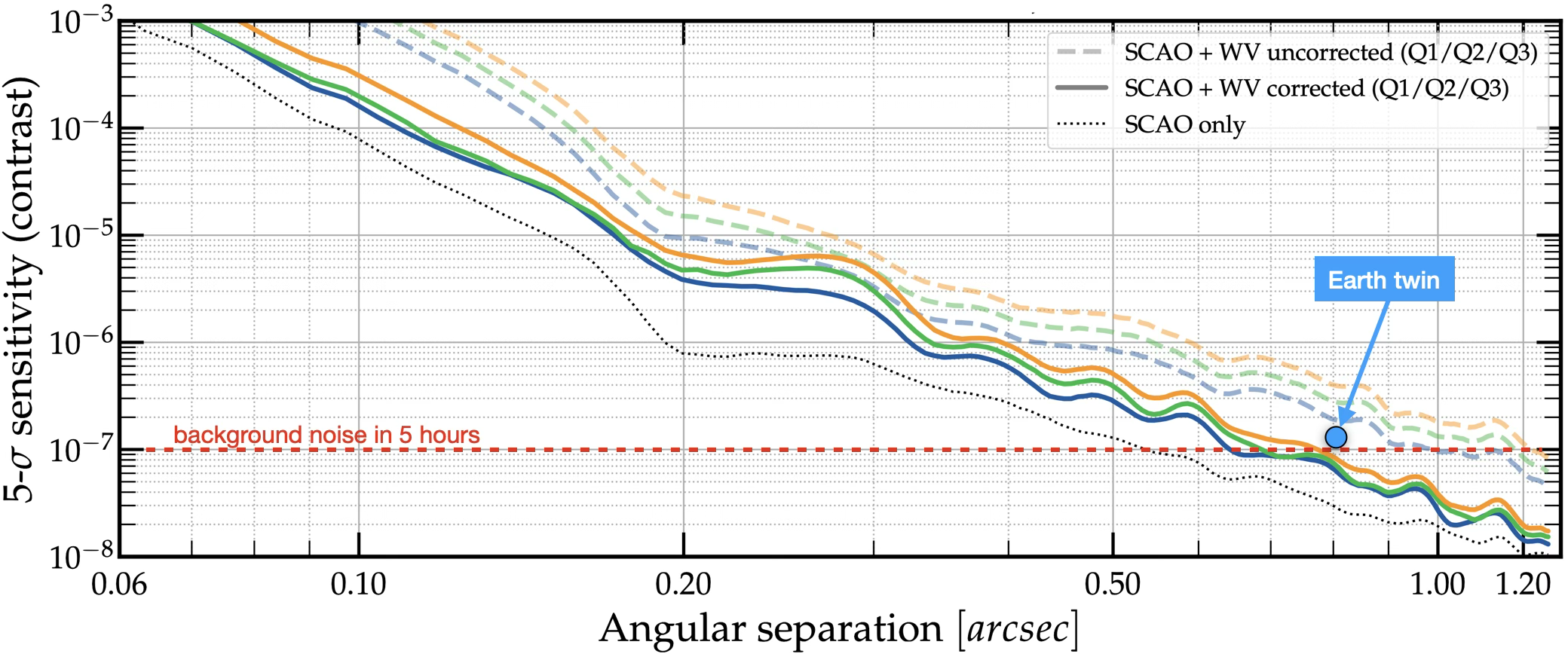}
  \end{tabular}
 \end{center}
 \caption[example] 
%>>>> use \label inside caption to get Fig. number with \ref{}
 { \label{fig:alphaCen} Influence of WV seeing on N-band sensitivity limits in terms of contrast, showing the cases of no WV seeing (dotted), uncorrected WV seeing (coloured dashed lines), and 10-Hz closed loop WV seeing correction on the first 20 Zernike modes (coloured solid lines) for various WV seeing quartiles. These simulations do not take into account any additional source of performance degradation beyond the wavefront errors from SCAO and WVS, such as chromatic beam wander and amplitude aberrations, which are beyond the scope of this study (see Refs.~\citenum{Carlomagno20,Delacroix22,Orban26} for a more thorough discussion of HCI performance in METIS).}
\end{figure}

As a last step, we focus on the case of alpha Cen, which is the highest profile target for N-band coronagraphy with METIS, as it provides the highest chances to detect and study a temperate rocky planet\cite{Bowens21}. In Fig.~\ref{fig:alphaCen}, we show the post-ADI sensitivity in terms of contrast based on a simulation where photon noise was not included to better highlight the effect of WVS on speckle noise. The thin black dotted curve shows the expected sensitivity in presence of only SCAO residuals for a 1-h ADI sequence on alpha Cen in the N2 band, while the coloured lines show how this sensitivity degrades in presence of WVS without correction (dashed lines) or with closed-loop focal-plane wavefront control (solid lines). In these simulations, we control the first 20 Zernike modes at a loop frequency of 10~Hz, which is in line with the sensitivity of focal-plane wavefront sensing described in Ref.~\citenum{Orban24}. We show the background-limited sensitivity in 5 hours of integration for comparison, as well as the expected brightness of an Earth twin around alpha Cen. Figure~\ref{fig:alphaCen} shows that controlling WVS should provide a gain ranging from a factor 3 to 5 in terms of sensitivity (depending on WVS conditions), and may have a decisive impact on the detectability of an Earth twin around alpha Cen.

\section{CONCLUSIONS AND PERSPECTIVES}
\label{sec:conclusion}  % \label{} allows reference to this section

In this study we have shown how, if not corrected, water vapour seeing (WVS) will drive the N-band performance of high-contrast imaging on ELT/METIS. We have shown that the harmful effect of WVS can be partly mitigated by closed-loop focal-plane wavefront control, down to a level that could still enable the detection of an Earth twin around alpha Cen with closed-loop control of 20 Zernike modes at 10~Hz. Further improving the performance of N-band coronagraphic observations would require to control more modes and/or implement a faster control loop. While the latter will eventually be limited by the frame rate of the METIS IMG detectors and the lag in the METIS data stream, controlling more modes could be considered, especially for bright stars like alpha Cen. Preliminary results from a more advanced control strategy based on reinforcement learning suggests that controlling 55 modes is within reach\cite{Nousiainen26,Taskin26}.

\acknowledgments % equivalent to \section*{ACKNOWLEDGMENTS}       
 
O.A.\ is a Research Director of the Fonds de la Recherche Scientifique – FNRS. This study received funding from the European Research Council (ERC) under the European Union's Horizon 2020 research and innovation programme (grant agreement No 819155), and from the Innovation in Science Pursuit for Inspired Research (INSPIRE) Fellowship, Department of Science and Technology, Government of India.

% References
\bibliography{report} % bibliography data in report.bib
\bibliographystyle{spiebib} % makes bibtex use spiebib.bst

\end{document}